\documentstyle[12pt]{article}
\def\journal{\topmargin .3in	\oddsidemargin .5in
	\headheight 0pt	\headsep 0pt
	\textwidth 5.625in % 1.2 preprint size  %6.5in
	\textheight 8.25in % 1.2 preprint size 9in
	\marginparwidth 1.5in
	\parindent 2em
	\parskip .5ex plus .1ex		\jot = 1.5ex}
\journal
\def\baselinestretch{1.2}
\catcode`\@=11
\def\marginnote#1{}
\newcount\hour
\newcount\minute
\newtoks\amorpm
\hour=\time\divide\hour by60
\minute=\time{\multiply\hour by60 \global\advance\minute by-\hour}
\edef\standardtime{{\ifnum\hour<12 \global\amorpm={am}%
	\else\global\amorpm={pm}\advance\hour by-12 \fi
	\ifnum\hour=0 \hour=12 \fi
	\number\hour:\ifnum\minute<10 0\fi\number\minute\the\amorpm}}
\edef\militarytime{\number\hour:\ifnum\minute<10 0\fi\number\minute}
\def\draftlabel#1{{\@bsphack\if@filesw {\let\thepage\relax
   \xdef\@gtempa{\write\@auxout{\string
      \newlabel{#1}{{\@currentlabel}{\thepage}}}}}\@gtempa
   \if@nobreak \ifvmode\nobreak\fi\fi\fi\@esphack}
	\gdef\@eqnlabel{#1}}
\def\@eqnlabel{}
\def\@vacuum{}
\def\draftmarginnote#1{\marginpar{\raggedright\scriptsize\tt#1}}
\def\draft{\oddsidemargin -.5truein
	\def\@oddfoot{\sl preliminary draft \hfil
	\rm\thepage\hfil\sl\today\quad\militarytime}
	\let\@evenfoot\@oddfoot	\overfullrule 3pt
	\let\label=\draftlabel
	\let\marginnote=\draftmarginnote
   \def\@eqnnum{(\theequation)\rlap{\kern\marginparsep\tt\@eqnlabel}%
\global\let\@eqnlabel\@vacuum}  }
\def\preprint{\twocolumn\sloppy\flushbottom\parindent 2em
	\leftmargini 2em\leftmarginv .5em\leftmarginvi .5em
	\oddsidemargin -.5in	\evensidemargin -.5in
	\columnsep .4in	\footheight 0pt
	\textwidth 10in	\topmargin  -.4in
	\headheight 12pt \topskip .4in
	\textheight 7.1in \footskip 0pt
	\def\@oddhead{\thepage\hfil\addtocounter{page}{1}\thepage}
	\let\@evenhead\@oddhead	\def\@oddfoot{}	\def\@evenfoot{} }
\def\numberbysection{\@addtoreset{equation}{section}
	\def\theequation{\thesection.\arabic{equation}}}
\def\underline#1{\relax\ifmmode\@@underline#1\else
	$\@@underline{\hbox{#1}}$\relax\fi}
\def\titlepage{\@restonecolfalse\if@twocolumn\@restonecoltrue\onecolumn
     \else \newpage \fi \thispagestyle{empty}\c@page\z@
	\def\thefootnote{\fnsymbol{footnote}} }
\def\endtitlepage{\if@restonecol\twocolumn \else \newpage \fi
	\def\thefootnote{\arabic{footnote}}
	\setcounter{footnote}{0}}  %\c@footnote\z@ }
\catcode`@=12
\relax
\def\figcap{\section*{Figure Captions\markboth
	{FIGURECAPTIONS}{FIGURECAPTIONS}}\list
	{Figure \arabic{enumi}:\hfill}{\settowidth\labelwidth{Figure 999:}
	\leftmargin\labelwidth
	\advance\leftmargin\labelsep\usecounter{enumi}}}
 \relax
\def\tablecap{\section*{Table Captions\markboth
	{TABLECAPTIONS}{TABLECAPTIONS}}\list
	{Table \arabic{enumi}:\hfill}{\settowidth\labelwidth{Table 999:}
	\leftmargin\labelwidth
	\advance\leftmargin\labelsep\usecounter{enumi}}}
 \relax
\def\reflist{\section*{References\markboth
	{REFLIST}{REFLIST}}\list
	{[\arabic{enumi}]\hfill}{\settowidth\labelwidth{[999]}
	\leftmargin\labelwidth
	\advance\leftmargin\labelsep\usecounter{enumi}}}
 \relax
\makeatletter
\newcounter{pubctr}
\def\publist{\@ifnextchar[{\@publist}{\@@publist}}
\def\@publist[#1]{\list
	{[\arabic{pubctr}]\hfill}{\settowidth\labelwidth{[999]}
	\leftmargin\labelwidth
	\advance\leftmargin\labelsep
	\@nmbrlisttrue\def\@listctr{pubctr}
	\setcounter{pubctr}{#1}\addtocounter{pubctr}{-1}}}
\def\@@publist{\list
	{[\arabic{pubctr}]\hfill}{\settowidth\labelwidth{[999]}
	\leftmargin\labelwidth
	\advance\leftmargin\labelsep
	\@nmbrlisttrue\def\@listctr{pubctr}}}
 \relax
\makeatother
\catcode`\@=11
\def\section{\@startsection {section}{1}{0pt}{-3.5ex plus -1ex minus
 -.2ex}{2.3ex plus .2ex}{\raggedright\large\bf}}
\catcode`\@=12
\newskip\humongous \humongous=0pt plus 1000pt minus 1000pt

\newif\ifdtup

\def\oldreffmt#1{\rlap{[#1]} \hbox to 2\parindent{}}

\def\figfmt#1{\rlap{Figure {#1}} \hbox to 1in{}}

\def\beq{\begin{equation}}
\def\eeq{\end{equation}}

\def\bea{\begin{eqnarray}}
\def\eea{\end{eqnarray}}
\catcode`\@=11
\def\eqnarray{\stepcounter{equation}\let\@currentlabel=\theequation
\global\@eqnswtrue
\global\@eqcnt\z@\tabskip\@centering\let\\=\@eqncr
\gdef\@@fix{}\def\eqno##1{\gdef\@@fix{##1}}%
$$\halign to \displaywidth\bgroup\@eqnsel\hskip\@centering
  $\displaystyle\tabskip\z@{##}$&\global\@eqcnt\@ne
  \hskip 2\arraycolsep \hfil${##}$\hfil
  &\global\@eqcnt\tw@ \hskip 2\arraycolsep $\displaystyle\tabskip\z@{##}$\hfil
   \tabskip\@centering&\llap{##}\tabskip\z@\cr}

\def\@@eqncr{\let\@tempa\relax
    \ifcase\@eqcnt \def\@tempa{& & &}\or \def\@tempa{& &}
      \else \def\@tempa{&}\fi
     \@tempa \if@eqnsw\@eqnnum\stepcounter{equation}\else\@@fix\gdef\@@fix{}\fi
     \global\@eqnswtrue\global\@eqcnt\z@\cr}

\catcode`\@=12
\font\tenbifull=cmmib10 % bold math italic
\font\tenbimed=cmmib10 scaled 800
\font\tenbismall=cmmib10 scaled 666
\relax

\def\single {
                \renewcommand{\baselinestretch}{1}
                \large
                \normalsize
                }
\def\baselinestretch{1.5}
\def\thefootnote{\fnsymbol{footnote}}
\def\ref#1{$^{#1)}$}
\newcommand{\lsim}{\mbox{ \raisebox{-1.0ex}{$\stackrel{\textstyle <}
{\textstyle \sim}$ }}}

\single
\begin{document}
\begin{titlepage}
\begin{center}
\today     \hfill    LBL-34744 \\ % replace x's with LBL number

\vskip .5in

{\large \bf Leptogenesis in Supersymmetric Standard Model with Right-handed
Neutrino}
\footnote{This work was supported by the Director, Office of Energy
Research, Office of High Energy and Nuclear Physics, Division of High
Energy Physics of the U.S. Department of Energy under Contract
DE-AC03-76SF00098.}

\vskip .5in

Hitoshi Murayama\footnote{On leave of absence from {\it Department of
Physics, Tohoku University, Sendai, 980 Japan}}  \\%[.5in]

{\em Theoretical Physics Group,
     Lawrence Berkeley Laboratory\\
     University of California,
     Berkeley, California 94720}
\vskip .5in

T. Yanagida \\%[.5in]

{\em Department of Physics,
     Tohoku University\\
     Sendai, 980 Japan}
\end{center}

\vskip .5in

\begin{abstract}
%insert abstract here
We show that the leptogenesis is automatic within the supersymmetric
standard model if there exist right-handed neutrinos with mass less than
$H_{\it inf}$, the expansion rate during the inflation.
Its scalar component grows during the inflation due to
the quantum fluctuation, oscillates coherently after the inflation, and
its decay generates lepton asymmetry. The scenario is naturally embedded
into $SO(10)$ GUT without any modifications. We also discuss the case
with heavier right-handed neutrino.
\end{abstract}
\end{titlepage}

%THIS PAGE (PAGE ii) CONTAINS THE LBL DISCLAIMER
%TEXT SHOULD BEGIN ON NEXT PAGE (PAGE 1)
\renewcommand{\thepage}{\roman{page}}
\setcounter{page}{2}
\mbox{ }

\vskip 1in

\begin{center}
{\bf Disclaimer}
\end{center}

\vskip .2in

\begin{scriptsize}
\begin{quotation}
This document was prepared as an account of work sponsored by the United
States Government.  Neither the United States Government nor any agency
thereof, nor The Regents of the University of California, nor any of their
employees, makes any warranty, express or implied, or assumes any legal
liability or responsibility for the accuracy, completeness, or usefulness
of any information, apparatus, product, or process disclosed, or represents
that its use would not infringe privately owned rights.  Reference herein
to any specific commercial products process, or service by its trade name,
trademark, manufacturer, or otherwise, does not necessarily constitute or
imply its endorsement, recommendation, or favoring by the United States
Government or any agency thereof, or The Regents of the University of
California.  The views and opinions of authors expressed herein do not
necessarily state or reflect those of the United States Government or any
agency thereof of The Regents of the University of California and shall
not be used for advertising or product endorsement purposes.
\end{quotation}
\end{scriptsize}

\vskip 2in

\begin{center}
\begin{small}
{\it Lawrence Berkeley Laboratory is an equal opportunity employer.}
\end{small}
\end{center}

\newpage
\renewcommand{\thepage}{\arabic{page}}
\setcounter{page}{1}
%THIS IS PAGE 1 (INSERT TEXT OF REPORT HERE)

Baryogenesis \cite{Sakharov} has been the most interesting overlapping
topic in the particle physics and cosmology. The original idea of the
baryogenesis using the out-of-equilibrium decay of superheavy particles
in the grand unified theories (GUT) \cite{Yoshimura} is still the most
popular mechanism. Meanwhile, it was found that the generation of the
initial $B-L$ asymmetry is necessary rather than the baryon asymmetry
$B$ itself \cite{KRS}.\footnote{There are also attempts to generate
baryon asymmetry without having net $B-L$ asymmetry \cite{KRS,EW,eR}.}
It was in fact pointed out that the lepton asymmetry $L$ can be
converted to the baryon asymmetry through the anomalous electroweak
interactions, and the decay of the right-handed neutrino produces the
initial lepton asymmetry \cite{FY}. Furthermore, the existence of the
right-handed neutrino leads to non-vanishing neutrino masses
\cite{seesaw}, which can explain the solar neutrino deficit
(for a combined analysis of three experiments, see \cite{solar}) via the
MSW mechanism \cite{MSW}.

On the other hand, the supersymmetry has been regarded as an elegant
mechanism to protect the huge hierarchy against the radiative
corrections \cite{SUSY}.
Since there are many scalar fields in the supersymmetric standard model,
new possibilities may arise for the baryogenesis. Indeed, Affleck and
Dine \cite{AD} proposed a scenario of baryogenesis where the scalar
fields have large values along a flat direction at the end of the
inflation, and their coherent rotation carries baryon number.
Unfortunately, their original model does not create $B-L$
asymmetry.\footnote{The original model is viable when the decay of the
flat direction occurs at the temperature below the electroweak phase
transition.} Therefore, it is interesting to see what
scenarios are possible for leptogenesis under the assumption of the
supersymmetry and the existence of the right-handed neutrino.

The aim of this letter is to point out that the leptogenesis is
automatic in the supersymmetric standard model if there exists a
right-handed neutrino with mass $M$ less than $H_{\it inf}$, where
$H_{\it inf}$ is the expansion rate of the universe at the end of the
inflation.\footnote{There is an upperbound $H_{\it inf} < 2.7 \times
10^{14}$~GeV \cite{Salopek} using the COBE data \cite{COBE} where the
bound is saturated when the gravitational radiation dominates the
density fluctuation. One may have significantly smaller $H_{\it inf}$ if
the scalar fluctuation dominates \cite{Turner}, but most of the models
predict $H_{\it inf} \geq 10^{12}$~GeV \cite{Salopek}.}
During the inflation, the scalar component
of the right-handed neutrino developes a large amplitude due to the quantum
fluctuation \cite{quantum}, and it begins to oscillate after the
inflation. The decay of the coherent oscillation generates lepton
asymmetry, in an analogous manner as in Ref.~\cite{MSYY}. The reheating
temperature can be low enough to avoid the gravitino problem
\cite{gravitino}. Furthermore, the scenario is naturally embedded into
the $SO(10)$ GUT.

The case when the mass of the right-handed neutrino is heavier than
$H_{\it inf}$ is also discussed. In this case the direction $\tilde{L} =
H_u$ is flat up to $\tilde{L} = H_u \lsim (h^{-2} H_{\it inf}^2
M)^{1/3}$, which has a lepton-number
violating operator $\frac{h^2 m}{M} (\tilde{L} H_u)^2$. Here, $h$ is the
Yukawa coupling constant in the superpotential $W = h L H_u N$,
and $m$ is the supersymmetry breaking mass. If the scalar fields
grow along this direction during the inflation, then leptogenesis occurs
{\it \'a la} Affleck--Dine.

Our starting point is the supersymmetric standard model with three
generations of right-handed neutrinos.\footnote{At least two
right-handed neutrinos are required to have CP-violation in the
Yukawa coupling.} Let us assume that (at least) one of the right-handed
neutrinos, say that of the first generation, is lighter than $H_{\it inf}$.
The quantum fluctuation of the scalar field $\phi$ with mass $M$ is
\cite{quantum}
\begin{equation}
\langle \phi^2 \rangle
	\simeq \frac{3}{8 \pi^2} \frac{H_{\it inf}^4}{M^2} ,
\end{equation}
with a coherence length
\begin{equation}
l \sim H_{\it inf}^{-1} \exp \left( \frac{3 H_{\it inf}^2}{2 M^2} \right).
\end{equation}
The scalar field becomes coherent over super-horizon distances if $M <
H_{\it inf}$, which can be regarded as a classical constant field.
Therefore, the right-handed sneutrino of the first generation
$\tilde{N}$ becomes a coherent classical field if $M < H_{\it inf}$, and
can have a large value of $\tilde{N} \sim H_{\it inf}^2/M$ at the end of
the inflation.\footnote{Even for the gauge non-singlet scalar fields
$\phi$ with $D$-term potential, it is natural to expect $\phi \sim
H_{\it inf}$ during the inflation, though their coherence length is
short, $l \sim H_{\it inf}^{-1}$. Then the right-handed sneutrino of
third generation cannot acquire a large value due to its large coupling
to slepton and Higgs fields, even if it is lighter than $H_{\it inf}$.
The right-handed sneutrinos both of first and second generations can
grow during the inflation even when all the other fields are fluctuating
at $O(H_{\it inf})$. We discuss that of the first generation only for
simplicity.} During the reheating process, the expansion rate rapidly
decreases, and the right-handed sneutrino begins to oscillate around the
origin. The coherent oscillation decays into $L
\tilde{H}_u$ or $\tilde{L} H_u$ and their CP-conjugates when $t \simeq
\Gamma_N^{-1}$, producing the lepton number density
\begin{equation}
	n_L = \epsilon M |\tilde{N}|^2.
\end{equation}
Here $|\tilde{N}|$ is the amplitude of the coherent oscillation, and
$\epsilon$ is the CP-asymmetry in the sneutrino decay into leptons and
anti-leptons \cite{FY,MSYY}.

The fate of the generated lepton asymmetry depends on the lifetime of
the inflaton $\Gamma_\psi^{-1}$. Assuming the initial value of the
right-handed sneutrino $\tilde{N}_0$ is smaller than the Planck mass
$M_P$, the coherent oscillation of the sneutrino does not dominate the
energy density of the universe during the reheating process. Then there
are three possible cases: (1) $\Gamma_N > \Gamma_\psi$ so that
$\tilde{N}$ decays during the reheating process, (2) $\Gamma_N <
\Gamma_\psi < \Gamma_N (M_P / \tilde{N}_0)^4$ so that $\tilde{N}$
decays after the reheating, but without dominating the energy density of
the universe, and (3) $\Gamma_\psi > \Gamma_N (M_P / \tilde{N}_0)^4$ so
that $\tilde{N}$ dominates the universe after the reheating before its
decay.  The yield of the lepton asymmetry $Y_L = n_L/s$ is calculated as
\begin{enumerate}
\item $\displaystyle Y_L = \epsilon \frac{\tilde{N}_0^2}{M M_P}
		\left( \frac{\Gamma_\psi}{M_P} \right)^{1/2}$,
\item $\displaystyle Y_L = \epsilon \frac{\tilde{N}_0^2}{M M_P}
		\left( \frac{\Gamma_\psi}{M_P} \right)^{1/2}$,
\item $\displaystyle Y_L = \epsilon \frac{(M_P \Gamma_N)^{1/2}}{M}$,
\end{enumerate}
for each cases. The expressions for the temperature at the sneutrino
decay $T_{dN}$ are different for each cases, such as
\begin{enumerate}
\item $\displaystyle T_{dN} = (M_P^2 \Gamma_\psi \Gamma_N)^{1/4}$,
\item $\displaystyle T_{dN} = (M_P \Gamma_N)^{1/2}$,
\item $\displaystyle T_{dN} = (M_P \Gamma_N)^{1/2}$.
\end{enumerate}
For the produced lepton asymmetry to be converted to the baryon
asymmetry by the electroweak anomaly, $T_{dN}$ has to be larger than the
critical temperature of the electroweak phase transition. This puts
constraints on model parameters. For instance, take values like
$\epsilon = 10^{-3}$, $M = 10^{10}$~GeV, $h=10^{-5}$, $\Gamma_N = h^2
M/8\pi \simeq 0.1$~GeV, $m_\psi = 10^{13}$~GeV, $\Gamma_\psi =
m_\psi^3/M_P^2 \simeq 10$~GeV, and $\tilde{N}_0 \simeq
10^{16}$~GeV,\footnote{We will discuss later that $\tilde{N}$ can grow
up to the GUT-scale even in the $SO(10)$ model.} the case (2) is
realized, and we obtain $Y_L = 3 \times 10^{-10}$, and $T_{dN} =
10^9$~GeV. Note that the final baryon asymmetry $Y_B$ is related to the
initial lepton asymmetry by $Y_B = 0.35 Y_L$, which is consistent with
the value required from the nucleosynthesis, $Y_B = 0.6$--$1 \times
10^{-10}$. Furthermore, $T_{dN}$ is high enough so that the lepton
asymmetry is converted to baryon one while the reheating temperature
$T_{RH} \simeq \sqrt{\Gamma_\psi M_P} \simeq 10^{10}$~GeV is low enough
to avoid the gravitino problem \cite{gravitino}.

So far we discussed the supersymmetric standard model with right-handed
neutrinos. We are tempted to embed the whole scenario into $SO(10)$ GUT
since there the existence of the right-handed neutrino is natural. It
seems, however, not easy to implement the scenario into the $SO(10)$
GUT at the first sight, since $\tilde{N}$ acquires the $D$-term
potential due to
the $SO(10)$ gauge interactions. Suppose the $D$-term potential is $V
\sim g^2 |\tilde{N}|^4$, it has a self-mass $M_{\it self}^2 \sim
g^2 |\tilde{N}|^2$.  Then its maximum possible value is only
$|\tilde{N}|
\sim H_{\it inf}$, and its coherence length is short, $l \sim H_{\it
inf}^{-1}$. Thus it cannot be regarded as a coherent field over many
horizons.

Our crucial observation is that the $D$-term decouples from the
potential of $\tilde{N}$ as far as the value of $\tilde{N}$ is smaller
than the $SO(10)$
breaking scale. Then the whole scenario can be extended to $SO(10)$
GUT without modifications, despite the first glance.
We will show below with an explicit example that the scalar partner of
the Nambu--Goldstone boson automatically shifts to absorb the
non-vanishing $D$-term of $\tilde{N}$, and the $D$-term potential
decouples (for a general discussion on the decoupling of the
$D$-term potential, see Ref.~\cite{HLW} for example).

Suppose $SO(10)$ is broken down to $SU(5)$ by the condensation of the
$SU(5)$ singlet components in the $\phi({\bf 126})$ and $\bar{\phi}({\bf
126}^*)$ representations,\footnote{The discussions are not altered
qualitatively when the $SO(10)$ is broken by {\bf 16} representation
Higgs multiplet.
The only difference is that the right-handed neutrino mass comes from a
non-renormalizable interaction between Higgs {\bf 16} and matter {\bf 16}
multiplets.}
\begin{equation}
\langle \phi_{\bf 1} \rangle = \langle \bar{\phi}_{\bf 1} \rangle = v,
\end{equation}
where the subscript ${\bf 1}$ denotes the $SU(5)$ singlet components, and
vacuum expectation values of $\phi_{\bf 1}$, $\bar{\phi}_{\bf 1}$ are
assumed to take $v
\sim 10^{16}$~GeV for definiteness of the discussions. For example, we
assume the superpotential as
\begin{equation}
W = k ( \phi \bar{\phi} - v^2 ) \chi ,
\end{equation}
with a singlet superfield $\chi$.\footnote{It is necessary to arrange
the superpotential to get rid of unwanted massless fields. An example is
to introduce $\xi ({\bf 16})$ and $\bar{\xi} ({\bf 16}^*)$, and take a
superpotential $W = M \bar{\phi} \phi + M' \bar{\xi} \xi + g(\phi \xi
\xi + \bar{\phi} \bar{\xi} \bar{\xi})$. However, the precise form of
the superpotential is irrelevant to our discussions, and we took the
simplest possible form.} One also has to add a term to the
superpotential which gives the mass of the right-handed neutrino,
\begin{equation}
W_N = f \phi \psi \psi,
\end{equation}
where $\psi ({\bf 16})$ is the matter multiplet and $f$ is a coupling
constant. The right-handed neutrino mass is given by $M = 2 f v$. %To be
%consistent with our assumption $M < H_{\it inf}$, $f$ should be small, $f <
%10^{-3}$.
The $\phi_{\bf 1}$ and $\bar{\phi}_{\bf 1}$ fields are expanded around the
minimum,
\begin{eqnarray}
\phi_{\bf 1} &=& v + \frac{1}{2} ( \varphi + i \eta), \\
\bar{\phi}_{\bf 1} &=& v - \frac{1}{2} ( \varphi + i \eta) ,
\end{eqnarray}
where $\eta$ is the true Nambu--Goldstone field which is absorbed into
the gauge field, and $\varphi$ is its scalar partner. We will focus on
$\varphi$ hereafter.

During the inflation, all the fields with GUT-scale mass rapidly drop
into their stationary points, including $\phi$. On the contrary, the
right-handed sneutrino
field is light compared to the expansion rate, and we adopt the
adiabatic approximation to keep $\tilde{N}$ fixed, and solve for $\varphi$.
The potential in the $\varphi$ and $\tilde{N}$ space is
\begin{equation}
V = g^2 \frac{5}{16} ( 4 v \varphi + |\tilde{N}|^2)^2
	+ \frac{k^2}{16} \varphi^4
	+ M^2	|\tilde{N}|^2
	\left( \left( 1 + \frac{\varphi}{2v} \right)^2
		+ \left|\frac{\tilde{N}}{2v} \right|^2 \right),
\end{equation}
where the first term is the $D$-term potential, and we have used $M = 2 f
v$. Since $\varphi$ has a
mass $\simeq g v$ from the $D$-term, it rapidly drops into the minimum,
$\varphi = - |\tilde{N}|^2/4v + O(|\tilde{N}|^6)$, and it stays at this
minimum during the slow rolling of the $\tilde{N}$ as far as $|\tilde{N}|
< v$. Then the potential for $\tilde{N}$ alone reads
\begin{equation}
V = M^2 |\tilde{N}|^2 + \frac{M^2 |\tilde{N}|^6}{64 v^4} + O(|\tilde{N}|^8).
\end{equation}
Therefore, the potential remains
essentially the same as in the non-$SO(10)$ case up to $\tilde{N}
\sim v$.\footnote{The disappearance of the $|\tilde{N}|^4$ term is an
accident in this example. There appear terms like $\frac{M^2}{v^2}
|\tilde{N}|^4$ in general, but the conclusion remains the same.}
$\tilde{N}$ can grow up to the GUT-scale $v$ during the inflationary
epoch, and has a super-horizon size coherence length $l \sim H_{\it
inf}^{-1} \exp(3 H_{\it inf}^2 / 2 M)$.

It is an interesting question what occurs when the right-handed neutrino
is heavier. An intriguing possibility is that the right-handed sneutrino
itself is the inflaton \cite{MSYY}, which is possible when $M = H_{\it
inf} \simeq 10^{13}$~GeV. In this scenario, right-handed sneutrino has a
large value $\gg M_P$ at
the birth of the universe, and rolls slowly down the potential, thereby
driving the chaotic inflation. The reheating process itself produces lepton
asymmetry just as in the previous scenario. However, it is very
difficult to embed this scenario in the $SO(10)$ GUT, since the $D$-term
potential protects $\tilde{N}$ to have such a large value.

\setcounter{footnote}{0}

If the right-handed neutrino is further heavier, $M > H_{\it inf}$, it
rapidly drops into its stationary point during the inflation. On the
other hand, the direction $\tilde{L} = H_u$ is flat up to to $\tilde{L}
= H_u \lsim (h^{-2} H_{\it inf}^2 M)^{1/3}$ as shown below. Therefore,
the scalar fields may become coherent along this direction during the
inflation.\footnote{This direction is flat only when other fields such
as scalar top $\tilde{t}$ vanish. It might be problematic when
$\tilde{t} \sim H_{\it inf}$ during the inflation. It is, therefore, not
clear to us whether the scalar fields grow to a particular direction
during the inflation. This is a general problem inherent in the
Affleck--Dine scenario. We take conservatively $\tilde{L} = H_u
\simeq O(H_{\it inf})$ at the end of the inflation for later
estimation.}
The potential of $\tilde{L}$, $H_u$, and $\tilde{N}$ reads as
\begin{eqnarray}
V &=& \left| h \tilde{L} H_u + M \tilde{N} \right|^2
	+ \left|h \tilde{L} \tilde{N} \right|^2
	+ \left|h H_u \tilde{N} \right|^2
	+ \frac{g_Z^2}{8} ( |\tilde{L}|^2 - |H_u|^2)^2 \nonumber \\
& &	+ \left( A h \tilde{L} H_u \tilde{N} + \frac{1}{2} B M \tilde{N}^2
		+ c.c. \right)
	+ m_L^2 |\tilde{L}|^2 + m_H^2 |H_u|^2.
\end{eqnarray}
Here, the SUSY-breaking terms $A, B, m_L, m_H$ are all of the order of
the weak-scale, $h$ is the Yukawa coupling, $g_Z = e/\sin\theta_W
\cos\theta_W$, and we restricted ourselves to the neutral components in
$\tilde{L}$ and $H_u$. Since it is assumed $M > H_{\it inf}$, $\tilde{N}$
rapidly drops into the minimum
\begin{equation}
\tilde{N} = -\frac{ h \tilde{L} H_u}{M^2 + |h \tilde{L}|^2 + |h H_u|^2},
\end{equation}
while $\tilde{L}$ and $H_u$ can be treated adiabatically. We can expand
the potential in terms of $h \tilde{L}/M$ and $h H_u / M$, and then the
potential reads as
\begin{eqnarray}
V &=& \frac{h^4}{M^2} |\tilde{L} H_u|^2 (|\tilde{L}|^2 + |H_u|^2)
	+ \frac{g_Z^2}{8} ( |\tilde{L}|^2 - |H_u|^2)^2
		\nonumber \\
& &	- \left( \frac{h^2 (A-B)}{M} (\tilde{L} H_u)^2 + c.c. \right)
	+ m_L^2 |\tilde{L}|^2 + m_H^2 |H_u|^2.
\end{eqnarray}
There are two crucial points in this potential. One is that the
disappearance of the $|h \tilde{L} H_u|^2$ term in the potential which
might have destroyed the flatness if existent. The potential remains $V
\lsim H_{\it inf}^4$ up to $\tilde{L} \lsim (h^{-2} H_{\it inf}^2
M)^{1/3}$. Therefore, the potential is flat even for third generation
$\tilde{L}$. The other is that there automatically arises lepton-number
violating operator $\frac{h^2 (A-B)}{M} (\tilde{L}
H_u)^2$.\footnote{Note that the existence of the right-handed neutrino
is not a necessity in this scenario. The only requirements are that the
flat direction $\tilde{L} = H_u$ remains flat up to a large scale, and
there is a lepton-number violating operator $(\tilde{L} H_u)^2$.}
Therefore, this model is an ideal and so-far the simplest realization of
the Affleck--Dine mechanism.\footnote{The authors of Ref.~\cite{Olive}
also pointed out that the leptogenesis is possible {\it \'a la}
Affleck--Dine when the $R$-parity is explicitly broken and one uses an
effective operator which arises at the three-loop level.} We have checked
that the following parameters give the lepton asymmetry $Y_L \simeq
10^{-8}$ according to the estimation in
Ref.~\cite{Linde-on-AD}: $M = 10^{16}$~GeV, $\Gamma_\psi = 10$~GeV, $m =
10^3$~GeV, $h = 1$, and the initial value $\tilde{L} = H_u \simeq H_{\it
inf}$ with $O(1)$ phase factor.

In summary, we found that the leptogenesis is automatic within the
supersymmetric models if there exist a right-handed neutrino lighter
than $H_{\it inf}$. Its scalar component grows during
the inflation, oscillates coherently after the inflation, and its decay
generates lepton asymmetry. The reheating temperature can be low
(typically $T < 10^{10}$~GeV), which is welcome to avoid the gravitino
problem. Furthermore, the scenario is naturally embedded into $SO(10)$
GUT without any modifications. For heavier right-handed neutrino,
leptogenesis via Affleck--Dine mechanism may be possible, if the scalar
fields grow along the $\tilde{L} = H_u$ direction during the inflation.

\noindent {\bf ACKNOWLEDGMENTS}

This work was supported by the Director, Office of Energy
Research, Office of High Energy and Nuclear Physics, Division of High
Energy Physics of the U.S. Department of Energy under Contract
DE-AC03-76SF00098.

%\noindent {\bf REFERENCES}

\end{document}